\documentclass[aps,prb,10pt,amsmath,amssymb,twocolumn,longbibliography,superscriptaddress]{revtex4-2}
\usepackage{hyperref}
\usepackage{graphicx}
\usepackage{mathtools}
\usepackage{physics}

\def\be#1\ee{\begin{align}#1\end{align}}

\begin{document}

\title{Diverging shift current responses in the gapless limit of two-dimensional systems}

\author{Hiroki Yoshida}
\affiliation{Department of Physics, Tokyo Institute of Technology, 2-12-1 Ookayama, Meguro-ku, Tokyo 152-8551, Japan}
\author{Shuichi Murakami}
\affiliation{Department of Physics, Tokyo Institute of Technology, 2-12-1 Ookayama, Meguro-ku, Tokyo 152-8551, Japan}

\date{\today}

\begin{abstract}
    The shift current responses of two-dimensional systems in the gapless limit are investigated. As the energy gap becomes smaller, the interband transition probability becomes larger at the band edges. We found the divergence of the shift current response in a manner proportional to the inverse of the gap size when the system becomes a two-dimensional Weyl semimetal in the gapless limit. This behavior is different from that for electric polarization, which has a jump across the gapless case. It means that the quantitative relationship between electric polarizations and shift currents is broken in this gapless limit.
\end{abstract}

\maketitle
\section{introduction}
\label{sec:intro}

The bulk photovoltaic effect (BPVE) is a DC current response to the uniformly irradiated light in non-centrosymmetric systems~\cite{kraut.vonbaltz1979a,belinicher.sturman1980,vonbaltz.kraut1981a,aversa.sipe1995a,sipe.shkrebtii2000,fridkin2001,fregoso.etal2017a,pusch.etal2023}. The BPVE has been actively investigated as it is a candidate for an alternative way to convert the light energy to the electric power. Since currents generated by the application of an external field oscillate in phase with that field, this BPVE is inherently a non-linear optical effect. Leading contributions to the BPVE are the second-order optical responses caused by the interband excitations of Bloch electrons. In the clean limit, generated second-order currents can be divided into two parts called injection currents~\cite{ahn.etal2020,ahn.etal2022} and shift currents~\cite{young.rappe2012,young.etal2012}. Injection currents and shift currents are caused by the change of the velocity and the position of electrons by the interband transition, respectively~\cite{sipe.shkrebtii2000}.

Injection and shift currents have been attracting much attention because they are related to the quantum geometry~\cite{Belinicher1982,sipe.shkrebtii2000,ahn.etal2022,hsu.etal2023b,morimoto.nagaosa2016,morimoto.nagaosa2016b,resta2011,ma.etal2021,morimoto.etal2023b}. It is known that the injection current has a direct connection with the Berry curvature~\cite{sipe.shkrebtii2000} and even quantization of the response occurs in Weyl semimetals~\cite{dejuan.etal2017a}. On the other hand, the shift current response is much more complicated and there are no direct connections between shift currents and geometric quantities such as Berry curvatures and Berry connections. Under some assumptions, it is found that the shift current is directly related with the Berry connection. As the integration of the Berry connection over the Brillouin zone is the electric polarization~\cite{Resta1992,King-Smith1993,Vanderbilt1993}, shift current can be described as a difference of the electric polarizations between conduction and valence bands~\cite{fregoso.etal2017a}. There are many previous studies that have investigated shift currents in topological materials~\cite{ahn.etal2020,chan.etal2017a,osterhoudt.etal2019,kim.etal2017,tan.rappe2016}, and quantization of shift currents by multi-gap topology has also been predicted~\cite{jankowski.slager2024}.

In this paper, we investigate the shift current response of two-dimensional systems when their band gap approaches zero. In our previous studies, we found that the electric polarization of a two-dimensional system has a jump when the system changes across a two-dimensional Weyl semimetal phase~\cite{Yoshida2023,yoshida.etal2023d}. Surprisingly, we found that this jump is solely determined by the Weyl dipole, described by the relative positions and monopole charges of Weyl points in the reciprocal space. Since the shift current is related with the electric polarization, we naively expect a jump or other singular behaviors of the shift current response across the two-dimensional Weyl semimetal phase. As a result, in the present paper we found that the shift current at the band edge diverges, in inverse proportion to the band gap in the limit. In the gapless limit, contributions from band edges solely determine the shift current, and when there are one conduction and one valence bands near the band edge, the coefficient of the divergence can be written using the first- and second-order derivatives of the two-band Hamiltonian with respect to the Bloch wave vector. These results indicate that the diverging behavior is a universal property of two-dimensional systems.

This paper is structured as follows. In Sec.~\ref{sec:gen}, we first develop a general theory of the shift current response at the band edge of two-dimensional systems in the gapless limit. We mainly study the case when the systems become Weyl semimetals in this limit. We also study the cases with higher-order energy dispersions at the end of the section. Section~\ref{sec:model} deals with the two-dimensional tight-binding model of electrons on a hexagonal lattice and numerically confirm the diverging behavior of the shift current. In Sec.~\ref{sec:pol}, we examine our results based on the relationship between the electric polarization and the shift current. Finally, Sec.~\ref{sec:conclusion} is devoted to the conclusion. 

\section{general theory of shift current divergences in two-dimensional gapless limits}
\label{sec:gen}

We first review the theory of second-order optical responses. After introducing the shift current, we investigate the shift current response in a two-band system in the limit where the energy gap closes and the bands form a Dirac cone. Then, we comment on general multi-band systems with energy dispersions of higher orders in the wavevector. Throughout this paper, we focus on spinless and non-interacting two-dimensional systems at zero temperature for simplicity.

\subsection{Review of the shift current}
The dc current density induced by the second-order optical response can be written as
\be
    j_{\mathrm{dc}}^a = \sigma_{\mathrm{dc}}^{abc}(0;\omega,-\omega)E_b(\omega)E_c(-\omega),
\ee
where $\omega$ is the frequency of the incident light, $j_{\mathrm{dc}}^a$ is the dc-current in the $a$ direction, $\sigma_{\mathrm{dc}}^{abc}$ is the conductivity tensor, $\vb{E}(\omega)$ is the electric field of the incident light, and $a,b,c$ are the Cartesian coordinates. In the clean limit, the third-rank tensor $\sigma_{\mathrm{dc}}^{abc}$ can be divided into the injection $\sigma_{\mathrm{inj}}^{abc}$ and the shift currents $\sigma_{\mathrm{shift}}^{abc}$ as
\be
    \sigma_{\mathrm{dc}}^{abc}=\sigma_{\mathrm{inj}}^{abc}+\sigma_{\mathrm{shift}}^{abc}.
\ee
The injection current is caused by the change of velocities of electrons upon optical excitations and its response tensor $\sigma_{\mathrm{inj}}^{abc}$ is linearly proportional to the relaxation time. The shift current is caused by the change of the positions of electrons by the excitation and its response tensor $\sigma_{\mathrm{shift}}^{abc}$ is independent of the relaxation time. In this paper, we focus on non-magnetic systems with time-reversal symmetry. In this case, depending on the polarization of the irradiated light, we can distinguish the injection and shift currents. When we irradiate a circularly polarized light to non-magnetic systems, the second-order response is limited to the injection current, while for the linearly polarized light, we only get the shift current. Therefore, we here focus on a shift current response to linearly polarized light~\cite{ahn.etal2020,watanabe.yanase2021,holder.etal2020a,dai.rappe2023b}. Then, for the linearly polarized light, the tensor $\sigma_{\mathrm{shift}}^{abb}$ can be explicitly written as~\cite{sipe.shkrebtii2000,morimoto.nagaosa2016b,kraut.vonbaltz1979a,vonbaltz.kraut1981a,barik.sau2020}
\be
    \sigma_{\mathrm{shift}}^{abb}& = -\frac{2\pi e^3}{\hbar^2}\int_{\vb{k}}\sum_{n,m}f_{nm}R^{ab}_{mn}\abs{r_{nm}^b}^2\delta(\omega_{mn}-\omega),\label{eq:shift-current-shiftvec}
\ee
where $-e\ (e>0)$ is the charge of an electron, $\int_{\vb{k}}\coloneqq\int_{\mathrm{BZ}}\frac{d^2k}{(2\pi)^2}$ is the integration over the two-dimensional Brillouin zone, $f_n\coloneqq f_F(\varepsilon_n)$ is the Fermi distribution function of the $n$-th band with energy $\varepsilon_n=\hbar\omega_n$, $f_{nm}\coloneqq f_n-f_m$ and $\hbar\omega_{nm}=\hbar\omega_n-\hbar\omega_m$ are the differences of the Fermi distribution functions and energies of the $n$-th and $m$-th bands, respectively, and $r_{nm}^b=\bra{n}i\partial_{k_b}\ket{m}$ is the $(n,\,m)$-element of the position operator. The quantity $R^{ab}_{mn} = -\partial_{k_a}\mathrm{arg}(r_{mn}^b)+r_{mm}^a-r_{nn}^a$ is called the shift vector. 
The above formula of the quantity $R^{ab}_{mn}$ involves derivatives of energy eigenvectors, and thus is not convenient for our study. Thus we use the formula for the shift current without the use of derivatives of energy eigenvectors in Ref.~\cite{cook.etal2017}. An equality for the off-diagonal elements of the position operator $r_{nm}^b=iv^b_{nm}/\omega_{mn}$ holds, where $v_{nm}^b = \bra{n}\partial H/\partial k_b\ket{m}/\hbar$ is the element of the velocity operator. Then, using these relations, $I_{mn}^{abb}\coloneqq R^{ab}_{mn}\abs{r_{nm}^b}^2$ can be expressed in terms of the velocity operators as
\be
     I_{mn}^{abb}&=\frac{1}{\varepsilon_{nm}^2}\Im\left[\frac{v_{nm}^a\qty(v_{mm}^b-v_{nn}^b)+\qty(a\leftrightarrow b)}{\varepsilon_{nm}}v_{mn}^b\right.\nonumber\\
    &\quad\left.+w_{nm}^{ab}v_{mn}^b-\sum_{p\neq n,m}\qty(\frac{v_{np}^bv_{pm}^a}{\varepsilon_{pm}}-\frac{v_{np}^av_{pm}^b}{\varepsilon_{np}})v_{mn}^b\right],\label{eq:integrand_gen}
\ee
where $\varepsilon_{mn}=\hbar \omega_{mn}$ and $w_{nm}^{ab}=\bra{n}\pdv[2]{H}{k_a}{k_b}\ket{m}$. The first two terms are direct excitation processes and the third term represents three-band processes, which vanishes for two-band models. Below, we consider a zero-temperature limit where $f_{nm}=-1$ if the $n$-th band is occupied and the $m$-th band is empty, $f_{nm}=0$ if both the $n$-th and $m$-th bands are occupied or empty.

\subsection{Shift current responses in the gapless limit}
For simplicity, we first confine our discussions to two-band gapped systems and calculate $I_{12}^{abb}$, where the band indices $m=1$ and $2$ denote the occupied and the unoccupied bands, respectively. Here we consider a system described by the Bloch Hamiltonian
\be
    H(\vb{k})=\mqty(M&g^*(\vb{k})\\g(\vb{k})&-M),\label{eq:2level_Hamiltonian}
\ee
where $g(\vb{k})$ is a complex function of $\vb{k}=\qty(k_x,k_y)$ and $M$ is a real parameter. Then, the energies of this system are 
\be
    \varepsilon_1 &= -\varepsilon,\\
    \varepsilon_2 &= +\varepsilon,\\
    \varepsilon &=\sqrt{\abs{g}^2+M^2}.
\ee
We assume that the complex function $g(\vb{k})$ becomes zero somewhere in the Brillouin zone. Then, the gap is given by $2\abs{M}$, and this system becomes gapless if and only if $g(\vb{k})=0$ and $M=0$. By a straightforward calculation of Eq.~\eqref{eq:integrand_gen}, $I_{12}^{abb}$ in the integrand can be written as
\be
    I_{12}^{abb}&=-\frac{M}{4\varepsilon^3}\Im\qty[\pdv{g^*}{k_b}\pdv[2]{g}{k_b}{k_a}-\frac{1}{2\varepsilon^2}\pdv{\abs{g}^2}{k_b}\pdv{g^*}{k_b}\pdv{g}{k_a}].\label{eq:two-level}
\ee
We can see that for all $\vb{k}$ such that $g(\vb{k})\neq0$, $I_{12}^{abb}(\vb{k})$ approaches $0$ in the gapless limit $M\to0$ due to the overall factor $M$ in Eq.~\eqref{eq:two-level}. On the other hand, at the band edge $\vb{k}_0$ such that $g(\vb{k}=\vb{k}_0)=0$, $\varepsilon$ becomes $\abs{M}$, $I_{12}^{abb}(\vb{k})$ has an overall factor $M^{-2}$ and the second term in the square bracket in Eq.~\eqref{eq:two-level} becomes zero. As a result, the shift current in the $M\to0$ limit is solely determined by the value of the first term in the square bracket in Eq.~\eqref{eq:two-level} evaluated at the band edge $\vb{k}_0$.

To calculate $\sigma_{\mathrm{shift}}^{abb}$, we need to consider the joint density of states (JDOS) part $\int_{\bf{k}}\delta\qty(\omega_{mn}-\omega)$. If the leading term of $g(\vb{k})$ around the band edge is proportional to $\vb{k}-\vb{k}_0$, the integration of the delta function in Eq.~\eqref{eq:shift-current-shiftvec} reduces to a term proportional to $\abs{M}$ as we see below. In order to carry out the integration with respect to the reciprocal vector, we need to change the variable of the delta function from $\omega$ to $\vb{k}$. Here, we follow the discussion in Ref.~\cite{cook.etal2017} and integrate around the band edge using a polar coordinate. Around the band edge $\vb{k}_0$ such that $g(\vb{k}_0)=0$, we can expand the function as $g(\vb{k})\approx \nabla_{\vb{k}} g\cdot(\vb{k}-\vb{k}_0)$, where the derivative is with respect to the reciprocal vectors. Then, we can approximate the energy $\varepsilon$ up to parabolic terms using $\delta\vb{k}\coloneqq\vb{k}-\vb{k}_0$ as
\be
    \varepsilon\approx \abs{M}+\frac{1}{2\abs{M}}\abs{\partial_{k_x}g(\vb{k}_0)\cdot\delta k_x+\partial_{k_y}g(\vb{k}_0)\cdot\delta k_y}^2.
\ee
We introduce a polar coordinate around $\vb{k}_0$ by variables $k$ and $\theta$ (see Appendix~\ref{app:higher order} and set $n=1$). Using these new variables, the JDOS at can be calculated as
\be
    \int_{\vb{k}}h(\vb{k})\delta(2\varepsilon-\omega) &\approx \int\frac{\mathrm{d}k\mathrm{d}\theta}{(2\pi)^2}\cdot\frac{h(\vb{k})\cdot 2\abs{M}k}{\abs{\Im[\partial_{k_x}g(\vb{k}_0)(\partial_{k_y}g(\vb{k}_0))^*]}}\nonumber\\
    &\qquad \times\frac{\delta\qty(k-\sqrt{(\omega-2\abs{M})\abs{M}})}{4k}\nonumber\\
    &= \frac{1}{2(2\pi)^2}\frac{\abs{M}}{\abs{\Im[\partial_{k_x}g(\vb{k}_0)(\partial_{k_y}g(\vb{k}_0))^*]}}\nonumber\\
    &\qquad \times\int \mathrm{d}\theta h(k,\theta)\eval_{k=\sqrt{(\omega-2\abs{M})\abs{M}}},\label{eq:JDOS_2band}
\ee
where $h(\vb{k})$ is an arbitrary function. As shown above, $h(\vb{k})$ for the calculation of the shift conductivity $\sigma_{\mathrm{shift}}^{abb}$ is $I_{12}^{abb}$, which is non-zero only at $k=0$. Then, combining Eqs.~\eqref{eq:two-level} and \eqref{eq:JDOS_2band}, in the gapless limit $M\to0$, we get
\be
    \sigma_{\mathrm{shift}}^{abb}(\omega)\eval_{\omega=2M} \sim \frac{e^3}{4\hbar^2}\sum_{g(\vb{k})=0}\frac{\Im\qty[\pdv{g^*}{k_b}\pdv[2]{g}{k_b}{k_a}]}{\abs{\Im[\partial_{k_x}g(\partial_{k_y}g)^*]}}\cdot\frac{1}{2M}.\label{eq:band_edge_gen}
\ee
On the other hand, contributions from points other than the band edge $(\omega>2M)$, the shift conductivity tensor becomes zero in the $M\to0$ limit. This is one of our main results. We can see that the shift current response for the light corresponding to the band edge is inversely proportional to the gap size and is divergent in the long-wavelength limit, or equivalently, the gapless limit. The coefficient of $1/M$ includes the first- and second-order derivatives of $g(\vb{k})$. These are generally non-zero in realistic systems.

This divergence is different from the low-frequency divergence proportional to $1/\omega^{4-d}$, where $d$ is the dimension of the Weyl semimetals discussed in Ref.~\cite{ahn.etal2020}. The low-frequency divergence discussed in Ref.~\cite{ahn.etal2020} is originated from the linear dispersion and it is present only if the Dirac cone is tilted and the system is in a state called type-II Weyl (Dirac) semimetal phase. The $1/\omega$ divergence discussed in this paper is present without the tilting of the Dirac cone.

\begin{figure}
    \begin{center}
        \includegraphics[width = 0.8\columnwidth]{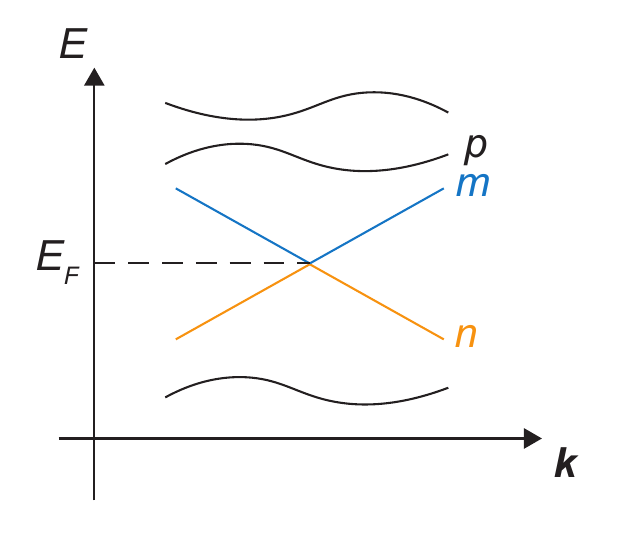}
        \caption{The schematic illustration of the multi-band system. The unoccupied $m$-th band and occupied $n$-th band form a Weyl point in the limit $M\to0$. The $p$-th band $(p\neq m,n)$ is not degenerate with these two bands.}
        \label{fig:multiband}
    \end{center}
\end{figure}

Next, we consider general multi-band systems. For systems with more than two bands, the third term in Eq.~\eqref{eq:integrand_gen} is nonzero. Here we analyze the $M$ dependency of this term. We consider three bands indexed as $m$, $n$, and $p$, where the $m$-th band is unoccupied and the $n$-th band is occupied and they form a Weyl point in the Weyl semimetal limit. The $p$-th band $(p\neq m,\,p\neq n)$ is not degenerate with these two bands (Fig.~\ref{fig:multiband}). First, $\varepsilon_{pm}$ approaches a finite value even in the Weyl semimetal limit because only the two bands $m$ and $n$ are degenerate in this limit. Likewise, terms of the form of $v_{np}^a$ are expected to approach a finite value in the limit $M\to0$. Next we discuss the terms that involve derivatives of eigenstates, such as $v_{np}^a=\bra{n}\ket{\partial_{k_a}p}/\varepsilon_{pn}$, $p\neq n$. As discussed above, $\varepsilon_{pn}$ approaches a finite value as $M\to0$. The amplitude of $\bra{n}\ket{\partial_{k_a}p}$ is proportional to the transition probability between two states $\ket{p}$ and $\ket{n}$. As the energies $\varepsilon_p$ and $\varepsilon_n$ are far apart, we expect the mixing of these two states does not behave singularly in the $M\to0$ limit. Therefore, the terms containing the sum over $p$ in Eq.~\eqref{eq:integrand_gen} are of the order of $M^0$. Thus even in the presence of these terms, the leading divergence in the two-dimensional Weyl semimetal limit is still $1/M$.

To end this section, we consider a higher-order band degeneracy such as quadratic band degeneracy as a comparison with a Weyl point. Suppose that $g(\vb{k})=(c_1\delta k_x+c_2\delta k_y)^n+\order{\abs{\delta\vb{k}}^{n+1}}\quad (c_1,c_2\in \mathbb{C},\,n\in\mathbb{N})$ around the gap minimum. Even in this case, contributions from $\vb{k}$-points where $g(\vb{k})\neq0$ goes to zero in the $M\to0$ limit as the integrand in Eq.~\eqref{eq:integrand_gen} has an overall factor $M$. For $\vb{k}$-points around $g(\vb{k})=0$, we can still carry out the integration of the JDOS by using a polar coordinate. Then, by similar arguments with the case of $n=1$ (for the details of the calculation, see Appendix~\ref{app:higher order}), we get
\be
    \sigma_{\mathrm{shift}}(\omega)\sim \frac{\mathrm{sgn}(M)}{\abs{M}^{2-\frac{1}{n}}\qty(\omega-2\abs{M})^{1-\frac{1}{n}}}\quad (\omega\gtrapprox2\abs{M}),
\ee
for $\omega$ slightly larger than $2\abs{M}$. When $n>1$, the shift current response is divergent at the frequency of the band edge even with a nonzero value of $M$. The divergent term $\abs{\omega-2\abs{M}}^{\frac{1}{n}-1}$ comes from the JDOS and we can see that the $1/M$ divergence discussed above is unique to the Weyl point with linear energy dispersion ($n=1$), where the JDOS is not divergent at the band edge.

\section{tight-binding model calculation}
\label{sec:model}

\begin{figure}
    \begin{center}
        \includegraphics[width = \columnwidth]{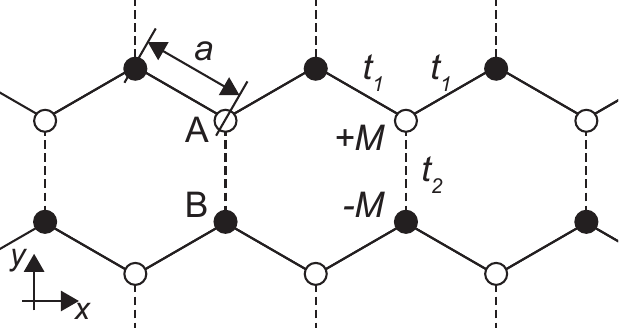}
        \caption{The tight-binding model on a two-dimensional hexagonal lattice. Anisotropic nearest-neighbor hoppings are parametrized by $t_1$ for the solid lines and $t_2$ for the dashed lines. The on-site potentials at the sites in the sublattices A and B are $+M$ and $-M$, respectively. The lattice constant is given by $a$. This system has an inversion symmetry if $M=0$ and has a $C_3$-rotational symmetry if $t_1=t_2$.}
        \label{fig:hexagonal_lattice}
    \end{center}
\end{figure}

In this section, we numerically calculate $\sigma_{\mathrm{shift}}^{abb}(\omega)$ for a two-dimensional tight-binding model of electrons on an anisotropic hexagonal lattice (Fig.~\ref{fig:hexagonal_lattice}) and confirm the divergent behavior of the shift current response in the gapless limit.

The tight-binding Hamiltonian of the system is given by
\be
    \hat{\mathcal{H}}&=\sum_{\langle i,j\rangle}t_{ij}\hat{c}_i^{\dagger}\hat{c_j}+\sum_i M_{i}\hat{c}_i^{\dagger}\hat{c}_i,\label{eq:TBH}
\ee
where the first sum is over all the pairs of neighboring sites. The parameter $t_{ij}\ (>0)$ describes the nearest-neighbor hoppings between neighboring sites $i$ and $j$. For the hoppings indicated by the solid lines in Fig.~\ref{fig:hexagonal_lattice}, we set $t_{ij}=t_1$ and for those by the dashed lines, we set $t_{ij}=t_2$. When $t_1\neq t_2$, this system lacks $C_3$-rotational symmetry. The real parameter $M_i$ is the on-site potential and takes the values $+M$ and $-M$ on the sublattices A and B, respectively. When $M\neq0$, this term breaks inversion symmetry. Together with the $C_3$-rotational symmetry breaking, this lack of inversion symmetry allows non-zero shift current generation. After the Fourier transformation, the Bloch Hamiltonian of this system is given in the form of Eq.~\eqref{eq:2level_Hamiltonian} with
\be
    g(\vb{k}) = 2t_1\cos\qty(\frac{\sqrt{3}a}{2}k_x)e^{i\frac{a}{2}k_y}+t_2e^{-iak_y}.
\ee
This $g(\vb{k})$ becomes zero at $(k_x,k_y)=\qty(2\cos^{-1}\qty(t_2/(2t_1))/(\sqrt{3}a),2\pi/(3a))$ and $\qty(2\cos^{-1}\qty(-t_2/(2t_1))/(\sqrt{3}a),4\pi/(3a))$ if $\abs{t_2/t_1}\leq2$, where we set $0\leq\cos^{-1}x\leq\pi$. At these points, the bands have an energy gap of $2\abs{M}$. When $M=0$, two bands are degenerate at these $\vb{k}$-points with linear dispersion around them and this system is a two-dimensional Weyl semimetal. By substituting this function $g(\vb{k})$ into the general formula in Eq.~\eqref{eq:two-level}, we can calculate the shift current conductivity. For the numerical calculation, we approximated the delta function with a Lorenzian as $\delta(\omega) \approx 1/\pi\cdot\epsilon/(\epsilon^2+\omega^2)$, where $\epsilon$ is a positive infinitesimal.

\begin{figure}[t]
    \begin{center}
        \includegraphics[width = \columnwidth]{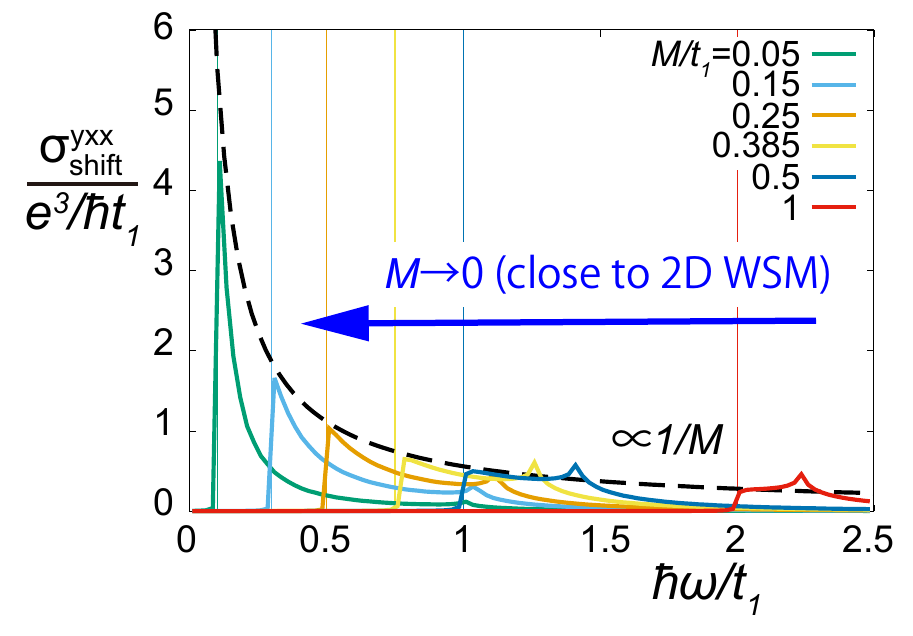}
        \caption{The shift conductivity $\sigma_{\mathrm{shift}}^{yxx}(\omega)$ for the tight-binding model \eqref{eq:TBH} for $M/t_1=0.05,0.15,0.25,0.385,0.5,1$. Vertical lines indicate the band edges $\omega=2M$. We set $t_2/t_1=0.5$. We set the Fermi energy $E_F$ to be $E_F=0$, which means that the bands are half-filled.}
        \label{fig:sigma_yxx}
    \end{center}
\end{figure}

The calculated $\sigma^{yxx}_{\mathrm{shift}}$ is plotted as a function of $\omega$ for several values of $M$ in Fig.~\ref{fig:sigma_yxx}. Solid curves with different colors indicate the calculated $\sigma^{yxx}_{\mathrm{shift}}$ of this system. The dashed curve indicates the expected $1/M$ divergence in Eq.~\eqref{eq:band_edge_gen}. The sizes of the band gap $2\abs{M}$ are indicated by vertical lines for each $M$ in the same color as $\sigma_{\mathrm{shift}}^{yxx}$. We can see that the shift current response is not divergent at band edges $\omega=2\abs{M}$ shown by the vertical lines for each value of $M$. On the other hand, when we make $M$ small, the shift current responses of the system at band edges diverges as $1/M$ and the curve shows a good agreement with the expected divergence from Eq.~\eqref{eq:band_edge_gen}.

Similar results can be obtained for $\sigma^{yyy}_{\mathrm{shift}}$. Note that since the system has a mirror symmetry with respect to the $y$-axis, $\sigma^{xaa}_{\mathrm{shift}}=0$ for $a=x,y$, and no currents are generated in the $x$-direction.

\section{Relationship with electric polarization}
\label{sec:pol}

The shift current is caused by the shift of positions of electrons along with the optical excitation and hence, related to the electric polarization of the material~\cite{fregoso.etal2017a}. In this section, we show that a simple relationship between electric polarizations and the shift current is lost in the two-dimensional Weyl semimetal limit in general. 

\begin{figure}[t]
    \begin{center}
        \includegraphics[width = 0.8\columnwidth]{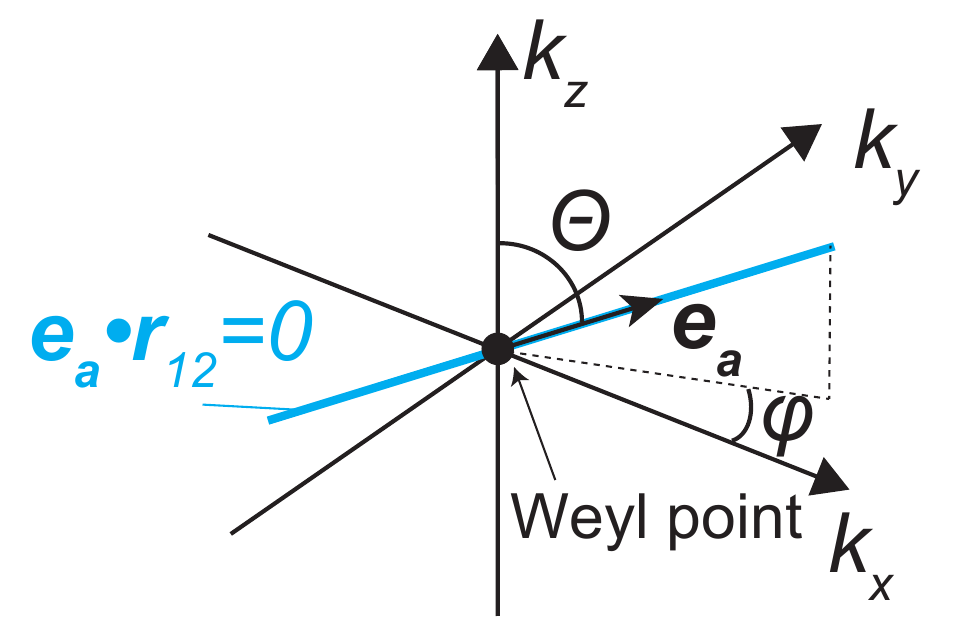}
        \caption{A Weyl point in the reciprocal space. Along the blue line passing the Weyl point in the direction of a unit vector $\vb{e}_a$, that has an azimuthal angle $\varphi$ and a polar angle $\theta$, the off-diagonal element of the position operator $\vb{r}_{12}=i\bra{1}\nabla_{\vb{k}}\ket{2}$ has no radial components around the Weyl point.}
        \label{fig:dirac_cone}
    \end{center}
\end{figure}

Consider an effective Hamiltonian of a three-dimensional Weyl point
\be
    H_{\mathrm{eff}}=v\vb{k}\cdot\boldsymbol{\sigma},
\ee
where $v$ is a positive constant and $\boldsymbol{\sigma}=\qty(\sigma_x,\,\sigma_y,\,\sigma_z)$ are Pauli matrices. Then, eigenstates of this Hamiltonian can be written as
\be
    \ket{n}&=\frac{1}{\sqrt{2\abs{\vb{k}}}}\mqty(\eta\sqrt{\abs{\vb{k}}+\eta k_z}\\e^{i\beta}\sqrt{\abs{\vb{k}}-\eta k_z}),\label{eq:ket_twoband}
\ee
where $\eta =(-1)^n\ (n=1,2)$ and $\beta=\tan^{-1}\qty(k_y/k_x)$ is an azimuthal angle of $\vb{k}$ in a spherical coordinate. The corresponding energies are $\varepsilon_1=-v\abs{\vb{k}}$ and $\varepsilon_2=+v\abs{\vb{k}}$.

We first check the behavior of the off-diagonal element of the velocity operator $\vb{v}_{12}$ in a direction parallel to an arbitrary unit vector $\vb{e}_a$. Then, under the gauge choice of Eq.~\eqref{eq:ket_twoband}, $v_{12}^a\coloneqq\vb{v}_{12}\cdot\vb{e}_a$ is
\be
    v_{12}^a&=\frac{v}{\hbar}\left\{\pdv{k_x}{k_a}\qty(\cos\alpha\cos\beta-i\sin\beta)\right.\nonumber\\
    &\quad +\pdv{k_y}{k_a}\qty(\cos\alpha\sin\beta+i\cos\beta)\nonumber\\
    &\quad\left.-\pdv{k_z}{k_a}\sin\alpha\right\},
\ee
where $\alpha=\tan^{-1}\qty(\sqrt{k_x^2+k_y^2}/k_z)$ is a polar angle of $\vb{k}$. As shown in Fig.~\ref{fig:dirac_cone}, let $\theta$ and $\varphi$ denote the polar and azimuthal angles of the vector $\vb{e}_a$ in the spherical coordinate. Then, using $\partial k_x/\partial k_a=\cos\varphi\sin\theta,\,\partial k_y/\partial k_a=\sin\varphi\sin\theta,$ and $\partial k_z/\partial k_a=\cos\theta$, the quantity $\abs{v_{12}^a}^2$ can be written using $\varphi$, $\theta$, and $\beta=\tan^{-1}\qty(k_y/k_x)$ as
\be
    \abs{v_{12}^a}^2&=\frac{v^2}{\hbar^2}\left\{\qty(\cos\alpha\sin\theta\cos\qty(\beta-\varphi)-\sin\alpha\cos\theta)^2\right.\nonumber\\
    &\qquad \left.+\sin^2\theta\sin^2\qty(\beta-\varphi)\right\}.
\ee
This quantity becomes zero if and only if $\qty(\varphi,\,\theta)=\qty(\beta,\,\alpha)$ or $\qty(\beta+\pi,\,\pi-\alpha)$, meaning that $\abs{v_{12}^a}=0$ on a line in $\vb{k}$-space crossing the Weyl point and parallel to the direction of $\vb{e}_a$ (blue line in Fig.~\ref{fig:dirac_cone}). This result is independent of the choice of gauges as $\abs{v_{12}^a}$ is a gauge-independent quantity. Since $r_{12}^a=iv_{12}^a/(2\varepsilon)$, $r_{12}^a$ also becomes zero on the line crossing the Weyl point and parallel to the $\vb{e}_a$.

Now, we discuss whether the shift current and the electric polarization behave similarly. In the Ref.~\cite{fregoso.etal2017a}, it is shown that if $r_{12}^a(\vb{k})$ can be approximated to be constant as a function of $\vb{k}$; then $\abs{r_{12}^a}^2$ in Eq.~\eqref{eq:shift-current-shiftvec} can be taken out of the integral, leading to the quantitative relationship between the shift current conductivities and the difference of electric polarizations as
\be
    \int\mathrm{d}\omega \sigma_{\mathrm{shift}}^{aaa}\approx -\frac{\pi e^2\abs{r^a_{12}}^2}{\hbar^2}\qty(P^a_2-P^a_1).
\ee
This relationship is consistent with the understanding of the shift current as a shift of the positions of electrons via optical excitations. In our setting, however, the element $\abs{r_{12}^a}$ goes to zero around the Weyl point in the two-dimensional Weyl semimetal limit. Hence, the assumption of the argument above is not valid in this case and the quantitative relationship between the shift conductivity and the electric polarization is lost in the two-dimensional Weyl semimetal limit.

\section{conclusion}
\label{sec:conclusion}

In this paper, we investigated the shift current response of two-dimensional systems when the system approaches a gapless limit and found a divergence of the shift conductivity when the system in the gapless limit is a Weyl semimetal. This work is motivated by the authors' previous findings that by a change of the system across a two-dimensional Weyl semimetal phase, the electric polarization has a discrete jump. By noting the quantitative relationships between the shift conductivity and electric polarizations in the previous works, we expected a jump of the shift conductivity across a two-dimensional Weyl semimetal phase.

The calculation of the shift conductivity around the band edge can be divided into the geometric part and the joint density of states. For the direct excitation of electrons from one band to another band, the geometric part is proportional to $M^{-2}$, where $M$ parameterizes the size of the band gap. As the joint density of state part is proportional to $M$ in the Weyl semimetal limit, the contribution from this direct excitation to the shift conductivity is proportional to $M^{-1}$ and diverges as we take the Weyl semimetal limit. We found that the coefficient of this divergence is given by the first- and second-order derivatives of the Hamiltonian with respect to the Bloch wavevector and confirmed its validity by numerical calculations. For multi-band systems, indirect excitation processes via transitions to another band are involved, making the divergence coefficient more complex but still we can show that the contribution to the shift conductivity also diverges as $M^{-1}$. Taken together, the shift conductivity in the two-dimensional Weyl semimetal limit is divergent, in inverse proportion to the size of the gap.

This divergence crucially depends on the dimensionality and the dispersion of the energy degeneracy and hence, is unique to the two-dimensional Weyl semimetal limit. The unexpected finding of divergence rather than a jump in shift conductivity was due to the fact that the quantitative relationship between electric polarization and shift conductivity is lost in the Weyl semimetallic limit, and the intuitive picture of the shift current as a change in the position of the center of charge can no longer be applied.

\begin{acknowledgments}
    This work is partly supported by Japan Society for the Promotion of Science (JSPS) KAKENHI Grant Numbers JP24KJ1109, JP22K18687, JP22H00108, and JP24H02231. HY is also supported by JST SPRING, Grant Number JPMJSP2106.
\end{acknowledgments}

\appendix
\section{Shift conductivity for energy degeneracies with higher order dispersions}
\label{app:higher order}
Here, we calculate the shift conductivity of a two-dimensional system in a gapless limit with higher-order energy dispersion in $\vb{k}$ discussed at the end of Sec.~\ref{sec:gen}. If an effective Hamiltonian is given in the same form as Eq.~\eqref{eq:2level_Hamiltonian}, we only need to consider contributions from band edges. Suppose that the off-diagonal component of an effective Hamiltonian $g(\vb{k})$ is written as
\be
    g(\vb{k})=\qty(c_1\delta k_x+c_2\delta k_y)^n+\mathcal{O}(\abs{\delta\vb{k}}^{n+1}),
\ee
where $c_1$ and $c_2$ are complex constants and $\delta k_x$ and $\delta k_y$ are the small displacements from the band edge. Then, the energy is approximated as
\be
    \varepsilon&\approx \abs{M}+\frac{1}{2\abs{M}}\abs{c_1\delta k_x+c_2\delta k_y}^{2n}\nonumber\\
    &=\abs{M}+\frac{1}{2\abs{M}}\qty(\abs{c_1}^2\delta k_x^2+2\Re(c_1c_2^*)\delta k_x\delta k_y+\abs{c_2}^2\delta k_y)^{n}\nonumber\\
    &=\abs{M}+\frac{1}{2\abs{M}}\qty(\mqty(\delta k_x&\delta k_y)\mqty(\abs{c_1}^2&\Re(c_1c_2^*)\\\Re(c_1c_2^*)&\abs{c_2}^2)\mqty(\delta k_x\\\delta k_y))^n\nonumber\\
    &=\abs{M}+\frac{1}{2\abs{M}}\qty(\mqty(\delta k_x'&\delta k_y')\mqty(\lambda_1&0\\0&\lambda_2)\mqty(\delta k_x'\\\delta k_y'))^n,
\ee
where $\lambda_{1,2}$ are positive numbers satisfying $\lambda_1+\lambda_2=\abs{c_1}^2+\abs{c_2}^2$ and $\lambda_1\lambda_2=\qty(\Im(c_1c_2^*))^2$. The original coordinate $(\delta k_x,\,\delta k_y)$ is transformed by an orthogonal matrix $O$ as $(\delta k_x',\,\delta k_y')^{\mathrm{T}}=O(\delta k_x,\,\delta k_y)^{\mathrm{T}}$. Then, we introduce a polar coordinate $(k,\,\theta)$ around the band edge as $\sqrt{\lambda_1}\delta k_x'=k\cos\theta$ and $\sqrt{\lambda_2}\delta k_y'=k\sin\theta$. Using these variables, we get
\be
    \varepsilon\approx \abs{M}+\frac{1}{2\abs{M}}k^{2n}.
\ee
In general, we can calculate the integration of our interest around the band edge as
\be
    &\phantom{=}\int\frac{\mathrm{d}\delta k_x\mathrm{d}\delta k_y}{(2\pi)^2}h(\vb{k})\delta\qty(2\varepsilon-\omega)\nonumber\\
    &=\int\frac{\mathrm{d}k\mathrm{d}\theta}{(2\pi)^2}h(\vb{k})\abs{\frac{\partial \qty(\delta k_x,\delta k_y)}{\partial\qty(k,\theta)}}\delta\qty(\frac{1}{\abs{M}}k^{2n}-\qty(\omega-2\abs{M}))\nonumber\\
    &=\int\frac{\mathrm{d}k\mathrm{d}\theta}{(2\pi)^2}\frac{h(\vb{k})\cdot k}{\sqrt{\lambda_1\lambda_2}}\frac{\delta\qty(k-\qty[(\omega-2\abs{M})\abs{M}]^{\frac{1}{2n}})}{\abs{\frac{2nk^{2n-1}}{\abs{M}}}}\nonumber\\
    &=\frac{1}{2n(2\pi)^2\abs{\Im(c_1c_2^*)}}\cdot\frac{\abs{M}^{\frac{1}{n}}}{\qty(\omega-2\abs{M})^{1-\frac{1}{n}}}\nonumber\\
    &\hspace{60pt}\times\int\mathrm{d}\theta h\qty(k,\,\theta)\eval_{k=\qty[(\omega-2\abs{M})\abs{M}]^{\frac{1}{2n}}}.
\ee
For the calculation of $\sigma_{\mathrm{shift}}^{abb}$, $h(\vb{k})$ is non-zero only at band edge in the limit $M\to0$ and thus, the $\theta$ integral yields $2\pi$. Combining this result with a $1/M^2$ dependence of other integrands of $\sigma_{\mathrm{shift}}^{abb}$, the $M$ dependence of $\sigma_{\mathrm{shift}}^{abb}$ is
\be
    \sigma_{\mathrm{shift}}^{abb}(\omega)\sim \frac{\mathrm{sgn}(M)}{\abs{M}^{2-\frac{1}{n}}\qty(\omega-2\abs{M})^{1-\frac{1}{n}}}.
\ee

\bibliography{ShiftCurrent.bib}

\end{document}